\documentclass[aps,prd,preprint,superscriptaddress,tightenlines,nofootinbib]{revtex4}

\usepackage{graphicx}% Include figure files
\usepackage{dcolumn}% Align table columns on decimal point
\usepackage{bm}% bold math
\usepackage{epsfig}
\def\etal{{\em et al.}}

\newcommand{\cleoiii}{CLEO\,III}
\newcommand{\cleoc}{CLEO-c}
\newcommand{\ccb}{\mbox{$c\overline{c}$}}

\newcommand{\mev   }{\mbox{\rm MeV}}

\newcommand{\mevcsq}{\mbox{\rm MeV/$c^2$}}
\newcommand{\gev   }{\mbox{\rm GeV}}

\newcommand{\invpb }{\mbox{\rm pb$^{-1}$}}

\newcommand{\decays}{\mbox{$\rightarrow$}}
\newcommand{\km}{\mbox{$K^-$}}

\newcommand{\pim}{\mbox{$\pi^-$}}
\newcommand{\pip}{\mbox{$\pi^+$}}
\newcommand{\piz}{\mbox{$\pi^0$}}
\newcommand{\phot}{\mbox{$\gamma$}}
\newcommand{\epem}{\mbox{$e^+ e^-$}}

\newcommand{\lz}{\mbox{$\Lambda$}}
\newcommand{\sigp}{\mbox{$\Sigma^+$}}
\newcommand{\sigz}{\mbox{$\Sigma^0$}}
\newcommand{\casmi}{\mbox{$\Xi^-$}}
\newcommand{\casz}{\mbox{$\Xi^0$}}
\newcommand{\casstrz}{\mbox{$\Xi^{*0}$}}
\newcommand{\omgmi}{\mbox{$\Omega^-$}}
\newcommand{\ppb}{\mbox{$p\overline{p}$}}
\newcommand{\lzlzb}{\mbox{$\Lambda\overline{\Lambda}$}}
\newcommand{\xmxmb}{\mbox{$\Xi^- \overline{\Xi^-}$}}
\newcommand{\xzxzb}{\mbox{$\Xi^0 \overline{\Xi^0}$}}
\newcommand{\omomb}{\mbox{$\Omega^- \overline{\Omega^-}$}}
\newcommand{\spspb}{\mbox{$\Sigma^+ \overline{\Sigma^+}$}}
\newcommand{\szszb}{\mbox{$\Sigma^0 \overline{\Sigma^0}$}}
\newcommand{\xsxsb}{\mbox{$\Xi^{*0} \overline{\Xi^{*0}}$}}

\begin{document}

%\preprint line(s) will be ignored for PRL/PRD
%\preprint{CLEO Draft 05-11A} % For paper draft CBX YY-NN -> Draft YY-NNA
%\preprint{CLEO CONF YY-NN}   % For conference papers
%\preprint{ICHEP ABSnnn}      % For conference papers
\preprint{CLNS 05/1918}       % for CLNS notes
\preprint{CLEO 05-10}         % for CLNS notes

\title{Branching fraction measurements of $\psi(2S)$ decay to baryon-antibaryon final states}

% Your author list goes here  REMOVE EVERYTHING to END INSERT and
% replace with your authorlist (ask cleoac).
%-------- INSERT ------------

%\author{(CLEO Collaboration)}     %FOR PRD_SPECIAL_CHANGEME
\author{T.~K.~Pedlar}
\affiliation{Luther College, Decorah, Iowa 52101}
\author{D.~Cronin-Hennessy}
\author{K.~Y.~Gao}
\author{D.~T.~Gong}
\author{J.~Hietala}
\author{Y.~Kubota}
\author{T.~Klein}
\author{B.~W.~Lang}
\author{S.~Z.~Li}
\author{R.~Poling}
\author{A.~W.~Scott}
\author{A.~Smith}
\affiliation{University of Minnesota, Minneapolis, Minnesota 55455}
% ...
\collaboration{CLEO Collaboration} %FOR PRL,CLNS
\noaffiliation

%-------- END INSERT ------------

\date{\today}
    
\begin{abstract}
Using 3.08 million $\psi(2S)$ decays observed in $e^+e^-$
collisions by the  CLEO detector, 
we present the results of a study of the $\psi(2S)$ decaying 
into baryon-antibaryon final states. 
We report the most precise measurements of the following eight modes: 
\ppb, \lzlzb, \xmxmb, \xzxzb\ (first observation), 
\spspb\ (first observation), and \szszb,
and place upper limits for the modes, \xsxsb\ and \omomb.

\end{abstract}

\maketitle

The study of $\psi(2S)$ production in \epem\, and its subsequent decay into 
two hadrons provides a  test of the predictive 
power of QCD~\cite{hel}, including information on gluon spin, 
quark distribution amplitudes in  baryon-antibaryon pairs, and total hadron 
helicity conservation. The decays are predicted to proceed via 
the annihilation of the constituent $c\overline{c}$ into three gluons or
a virtual photon. 
This model leads to the prediction that the ratio of the branching fraction 
into a specific final state to the branching fraction of the $J/\Psi$ into that
same state should be a constant value of approximately 13$\%$,
the corresponding ratio for the dilepton final state~\cite{rule}. 
This rule, which was previously refereed to as the ``12\% rule'',
is roughly obeyed for several channels, 
but fails for others~\cite{obey}.  
This Letter concentrates on the investigation of the two-body decays
of the $\psi(2S)$ into a baryon-antibaryon pair. 
Previous measurements~\cite{pdg} of the $\psi(2S)$ decaying into 
these final states have very large statistical 
uncertainties. 	   
The CLEO~\cite{cleoc} detector, with the advantage  
of good charged and neutral particle detection efficiencies,
together with secondary and tertiary 
vertex reconstruction, gives us the opportunity to make the most
precise measurements yet of these branching fractions. 

The data used in this analysis were collected at the CESR \epem\ storage ring,
which has been reconfigured to run near \ccb\ threshold by inserting 6
wigglers~\cite{cleoc}. Our analysis is based on 3.08$\times 10^6$ $\psi(2S)$ decays
which corresponds to the total integrated luminosity of 5.63 \invpb. 
Approximately half of the 
$\psi(2S)$ data (2.74 \invpb) were taken with the \cleoiii\ detector 
configuration~\cite{cleo}
and the remainder (2.89 \invpb) of the $\psi(2S)$ data, 
and all the continuum data 
(20.70 \invpb, $\sqrt{s} = 3.67~\gev$) were taken with the reconfigured 
\cleoc\ detector~\cite{cleoc}. 
We generated 10,000 Monte Carlo events using simulations of each of the two
detector configurations for each of the eight decay modes, using a
GEANT-based \cite{geant} detector modeling program. For all modes
we generated Monte Carlo samples with a 
flat distribution in $\cos\theta$, where 
$\theta$ is the angle of the decay products in the center-of-mass system
(in colliding beam experiments this angle is measured relative to the beam axis). 
For spin 1/2 baryons in the baryon octet, we then weighted the Monte Carlo samples 
with a $1+\cos^2\theta$ angular distribution, in agreement with the naive expectation.
Possible deviations from these angular distributions will be one source of systematic 
uncertainty in our measurements.

We begin by reconstructing the hyperons in the following decay modes
(branching fractions~\cite{pdg} are listed in parentheses):
\lz\decays $p$\pim (63.9$\%$), \sigp\decays$p$\piz (51.6$\%$),
\sigz\decays\lz\phot (100.0$\%$), \casmi\decays\lz\pim (99.9$\%$), 
\casz\decays\lz\piz (99.5$\%$), \casstrz\decays\casmi\pip (51.6$\%$),
and \omgmi\decays\lz\km (67.8$\%$). 
To discriminate between protons, kaons, pions, and electrons, 
we combined specific ionization ($dE/dx$) measured in 
the drift chamber~\cite{cleo} and log-likelihoods obtained from the RICH 
sub-detector~\cite{cleo}
to form a joint log-likelihood difference: 
${\cal L}(p-\pi)=L_{RICH}(p)-L_{RICH}(\pi)+\sigma^2_{dE/dx}(p)-\sigma^2_{dE/dx}(\pi)$, 
where the more negative ${\cal L}(p-\pi)$, the higher the likelihood that the particle 
is a proton compared to a pion. Our requirement on these quantities 
varies in value from mode to mode
depending upon background considerations.
Further details of this procedure may be found elsewhere~\cite{pkkpi}.
For protons in the \ppb\ and \sigz\ decay modes we require 
${\cal L}(p-\pi) < -9 $ and ${\cal L}(p-K) < -9$ (3 $\sigma$ separation), 
and for those in the 
\ppb\ mode we require additional criteria ${\cal L}(p-e) < -9$ 
as, in this case alone, electrons are a potentially significant background. 
For protons
that are the daughters of $\Lambda$ decays we make the looser requirements
of ${\cal L}(p-\pi) < 0 $ and ${\cal L}(p-K) < 0$, which are both 
very efficient. Kaons from $\Omega^-$
decays are strongly identified with ${\cal L}(K-\pi) < -9 $ 
and ${\cal L}(K-p) < -9$. Pions are only required to have energy loss 
measurements consistent with the pion identity.
Photon candidates were identified in the CsI calorimeter. 

\begin{figure}
\includegraphics*[width=3.0in]{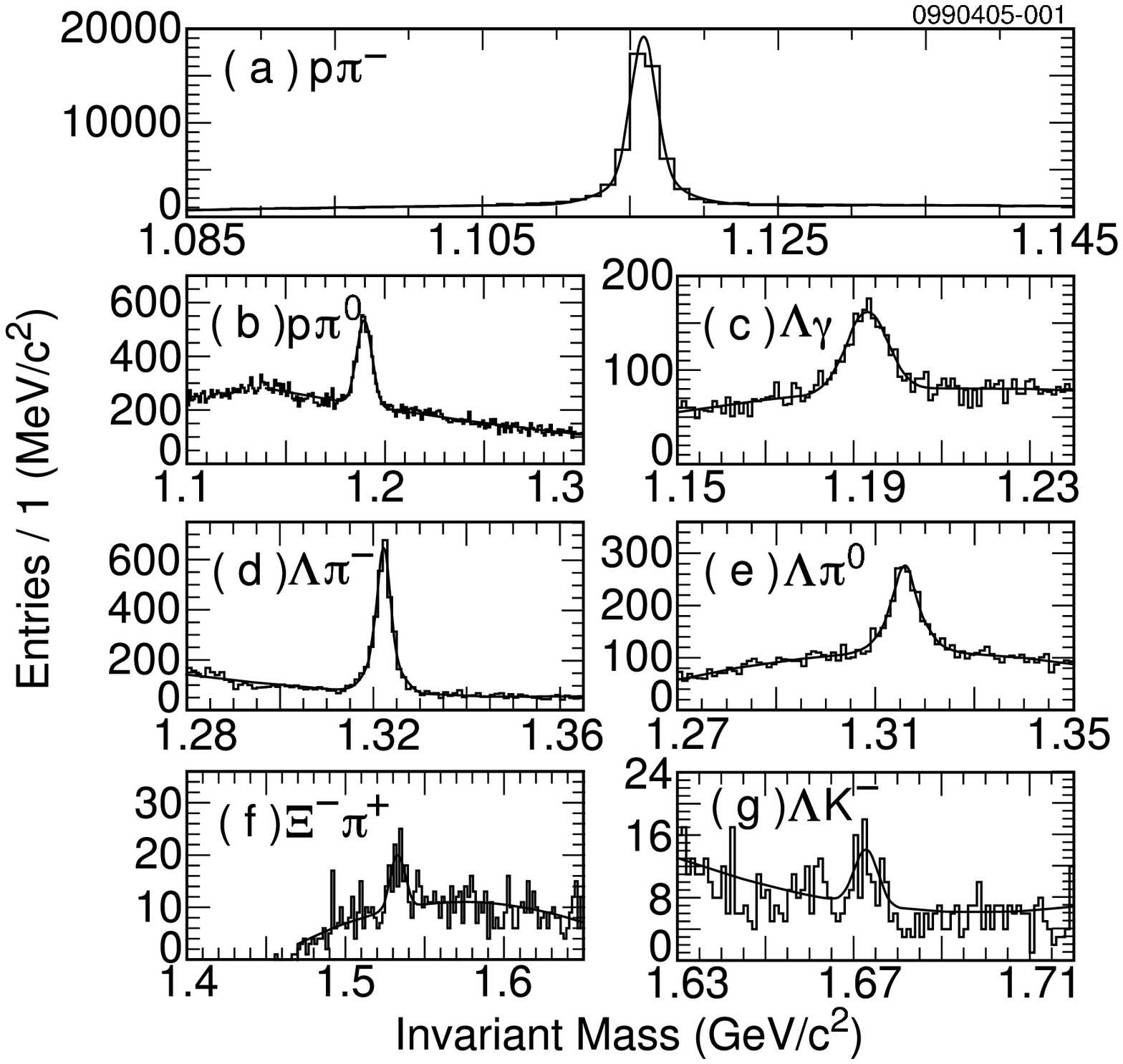}
\caption{Inclusive hyperon invariant mass distribution in $\psi(2S)$ data: 
(a) \lz\decays p\pim, (b) \sigp\decays p\piz, (c) \sigz\decays\phot\lz, 
(d) \casmi\decays\lz\pim, (e) \casz\decays\lz\piz, 
(f) \casstrz\decays\casmi\pip, (g) \omgmi\decays\lz\km.}
\label{fig:hyp}
\end{figure}

The analysis procedure for reconstructing \lz\ and \casmi\
closely follows that presented earlier~\cite{pkkpi}. 
The \omgmi\
reconstruction follows the steps of the \casmi\ reconstruction by 
replacing a pion by a kaon. \sigz\ hyperons are reconstructed
by combining a \lz\ with a photon and requiring the photon energy in the 
crystal calorimeter to be in excess of 50 \mev,  not matched to a charged track, 
and consistent in shape to that from a photon. 
\casz\ and \sigp\ hyperon reconstruction is complicated by the fact
that there is no direction information from the CsI photon clusters
for the \piz\ reconstruction.
A kinematic fit is made to the hypothesis that the parent hyperon started
at the beamspot, and decayed after a positive pathlength to a decay
vertex which is the origin of the $\pi^0 \to \gamma \gamma$ decay. A cut was
placed on the $\chi^2$ of the fit to this topology, which includes the fit to the
$\pi^0$ mass from the newly found hyperon decay vertex.
%We require this $\chi^2$ to be less than 20(10) for \casz (\sigp ) reconstruction.
The \casstrz\ candidate is reconstructed by vertexing
a positively charged pion candidate, consistent with coming from the beamspot, 
with the already found \casmi. 
All hyperon candidates were required to have vertices significantly 
separated from the beamspot,
with the flight distance of the \lz\, measured from the parent (\casz, \casmi,
or \omgmi) decay vertex, to be positive. 
In Figure~\ref{fig:hyp} we show the inclusive
hyperon yields in the $\psi(2S)$ data. 
We observed 48$\times10^3$ \lz\ candidate events in the
$\psi(2S)$ data as shown in Fig~\ref{fig:hyp}(a) with a fitted width ($\sigma$) 
found to be 1.53 \mevcsq, consistent with Monte Carlo estimates.  
The fitted yields (widths) in  Fig.~\ref{fig:hyp}(b) and (c) 
for \sigp\ and \sigz\ candidates are approximately 3$\times10^3$ (4.0 \mevcsq) 
and 1$\times10^3$ (4.7 \mevcsq),
respectively. Invariant mass distributions for the \casmi\ and \casz\ are shown in 
Fig.~\ref{fig:hyp}(d) and (e), respectively. The fit yielded approximately
2.7$\times10^3$ and 1.2$\times10^3$ for 
the signal events and 2.4 \mevcsq\ and 3.6 \mevcsq\ for widths for the \casmi\ and
\casz\ candidates, respectively. The fits in Fig.~\ref{fig:hyp}(f) and (g) yielded 
115\,(4.69 \mevcsq) and 47\,(2.69 \mevcsq) events\,(width) for the \casstrz\ and 
\omgmi\ spin 3/2 baryons.
Hyperon candidates within $3\sigma$ of their nominal masses 
are considered for further analysis.
We then combined the 4-momenta of these baryons with the  
corresponding 4-momenta of their charge conjugates and formed $\psi(2S)$ candidates.

We note that baryon-antibaryon pairs produced by $\psi(2S)$ decays have
a back-to-back topology, and each decaying baryon has exactly the beam energy.
Momentum conservation is imposed on the reconstructed
baryon-antibaryon pairs by demanding the vector sum of the total momentum  
in an event  $| \Sigma \vec{\bf p}| / E_{\rm beam}$ be less than 0.04. 
This eliminates background, via $\pi\pi J/\psi$ or $\phot\chi_{\rm cn}$
(n $=$ 0,1, and 2), which have extra tracks or showers. 

For each baryon-antibaryon candidate event, we calculated the scaled
visible energy, $E_{\rm vis}/\sqrt{s}$, where $E_{\rm vis}$ is the energy observed 
in an event and $\sqrt{s}$ is the center of mass energy. We define our 
signal region be $0.98 < E_{\rm vis}/\sqrt{s}<  1.02$, and two sideband
regions of 0.94-0.98 and 1.02-1.06 as representative of the 
combinatorial background. 

We also studied the continuum data to check for a possible contribution 
to our $\psi(2S)$ signals from this source. This was found to be non-negligible
for four of the decay modes as shown in Table~\ref{tab:result}. We multiplied the 
yield from the continuum data by a scaling factor which was calculated 
taking into account the differences in luminosity (0.2720), a $1/s^5$ correction (0.9572) for 
baryons~\cite{hel}, and the values of the
efficiencies in the \cleoiii\ and \cleoc\ detector configurations before 
subtracting it from the $\psi(2S)$ yields. This scaling factor was 0.2547 for 
the \ppb\ case and similar for the other modes.

Other possible
background sources are the daughters from the $J/\psi$ decay combined with
the charged or neutral transition pions which can produce possible 
cross-feed in the signal region.
To estimate this cross-feed  we generated corresponding Monte Carlo samples 
of 20,000 events and looked for events which passed our selection criteria 
for the modes concerned.

\begin{figure}
\includegraphics*[width=3.5in]{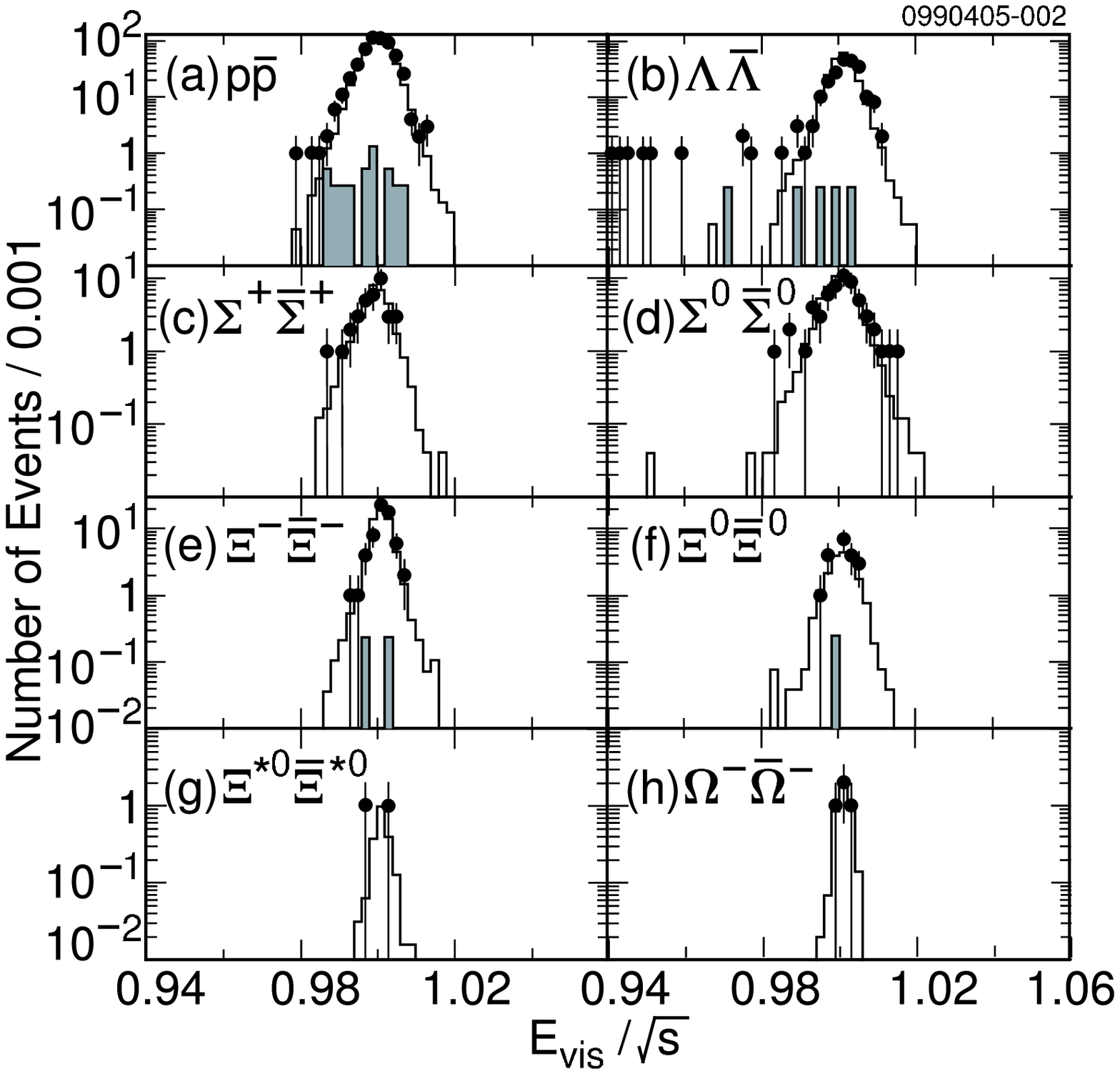}
\caption{Scaled energy ($E_{\rm vis}/\sqrt{s}$) distribution for eight $\psi(2S)$ 
decay modes: (a) \ppb, (b) \lzlzb, (c) \spspb, (d) \szszb, (e) \xmxmb, (f) \xzxzb,
(g) \xzxzb, and (h) \omomb. Signal Monte Carlo, data, and scaled continuum plots 
are shown in solid line, error bar, and filled histograms, respectively.}
\label{fig:data}
\end{figure}

Figure~\ref{fig:data} shows the scaled energy distribution for each of the decay modes. In
all cases, clear signals are seen, with widths as expected from Monte Carlo 
simulation, and very little combinatorial background in the sideband regions.
In each mode we calculate the number of $\psi(2S)$ decays 
to each final state (``signal yield'')
$N_S = S_{\psi(2S)}-B_{\psi(2S)}-S_{{\rm x}f}-f_s\cdot B_{c}$ (where $S_{\psi(2S)}$ is 
the total number of events in the signal region, $B_{\psi(2S)}$ events in the scaled sidebands,
 $S_{{\rm x}f}$ is the contribution from the cross-feeds, and $f_s\cdot B_{c}$ is 
the scaled continuum contribution with scaled sidebands subtracted). 
These yields are shown in Table~\ref{tab:result}. The efficiencies,
calculated using a weighted average of results from \cleoiii\ and \cleoc\   
detector simulations, are also shown.

\begin{small}
\begin{table*}[hp]
\begin{center}
\caption{Branching ratios of $\psi(2S)$ decaying to baryon-antibaryon pairs.  
 The second, third, fourth, and fifth columns
show the yields in the signal region, scaled sidebands, scaled continuum, and
cross-feeds, respectively. The sixth and seventh columns tabulate the efficiencies
and the measured branching fractions, respectively. We used Poisson statistics in 
evaluating the upper limits on the \xsxsb\ and \omomb\ branching fractions $@$90CL.
The eighth column shows the ratio of 
$\psi(2S)$ to $J/\psi$ decays with statistical and systematic uncertainties 
added in quadrature. The last column shows the background subtracted continuum cross-section and the
systematic error includes 7$\%$ due to initial and final state radiation effects and  
1$\%$ luminosity uncertainty (replacing the uncertainty on the number of $\psi(2S)$ decays).}
\label{tab:result}
\begin{tabular}{l c c c c c c c c} \hline
Modes    &$S_{\psi(2S)}$&$B_{\psi(2S)}$&$f_S\cdot B_c$&$B_{{\rm x}f}$& $\epsilon$&{${\cal B}(10^{-4})$} &Q($\%$)&$\sigma_{cont}(pb)$\\ \hline
\ppb     &557           & 0.5          &4.06 &0       & 66.6\% &2.87$\pm$0.12$\pm$0.15&13.6$\pm$1.1 & 1.5$\pm$0.37$\pm$0.13 \\
\lzlzb   &208           & 4.5          &0.86 &0       & 20.1\% &3.28$\pm$0.23$\pm$0.25&25.2$\pm$3.5 & $<$2.0 $@$90 CL\\
\spspb   &35            & 0.5          &0 &0.3        & 4.1\% &2.57$\pm$0.44$\pm$0.68& - &   -       \\
\szszb   &58            & 0            &0    &0       & 7.2\% &2.63$\pm$0.35$\pm$0.21&20.7$\pm$4.2 & -\\
\xmxmb   &63            & 0            &0.46 &0       & 8.6\% &2.38$\pm$0.30$\pm$0.21&13.2$\pm$2.2& $<$3.5 $@$90 CL \\
\xzxzb   &19            & 0            &0.49 &0       & 2.4\% &2.75$\pm$0.64$\pm$0.61& & $<$14 $@$90 CL  \\
\xsxsb   &2             & 0            &0    &0.6     & 0.6\% &$0.72^{+1.48}_{-0.62}\pm0.10$& - & -\\ 
         &              &              &     &        &       &($<$3.2 $@$90 CL)& & \\ 
\omomb   &4             & 0            &0    &0       & 1.9\%&$0.70^{+0.55}_{-0.33}\pm0.10$& - & -\\ 
         &              &              &     &        &       &($<$1.6 $@$90 CL)& & \\ \hline 
\end{tabular}
\end{center}
\end{table*}
\end{small}

%systematics 

We evaluated the following systematic uncertainties to our measured
branching fractions:
3$\%$ uncertainty on the number of $\psi(2S)$ decays in our sample; 1$\%$
uncertainty in the simulation of our hardware trigger; 
1$\%$ uncertainty in the reconstruction in each charged track in the event; 
1$\%$ uncertainty for proton identification of tracks coming from the beamspot, 
and 2$\%$ for proton and kaon identification of tracks coming from the secondary 
vertices; 1$\%$ and 2$\%$ uncertainties for photon detection and 
\piz\ reconstruction, respectively; and 1$\%$ for background subtraction in 
\lzlzb\ mode. The detection efficiency also depends upon the angular 
distribution of the hyperons. We find a 10$\%$ change  in efficiency 
when we change from a flat distribution in
$\cos\theta$ to one of the form $1+\lambda\cos^2\theta$ (where $\lambda=1$). The
value of $\lambda$ may not be strictly 1 because of effects due to finite charm 
quark mass~\cite{model} and hadron mass effect from ${\cal O}(v^2)$ and higher twist
corrections to the effective QCD Lagrangian. We compare the detection efficiencies 
for spin one-half baryon generated samples
as discussed above with that obtained by the E835 collaboration~\cite{e835}
and assign the difference, 3.3$\%$, as the systematic uncertainty.
%We assign a 3.3$\%$ systematic uncertainty in our 
%efficiency for detecting spin one-half baryons with this angular distribution 
%and that as suggested by the E835 collaboration~\cite{e835}.
We assign a 10$\%$ uncertainty in angular distribution to spin 3/2 baryons, which 
covers all reasonable possibilities. 
The systematic uncertainty in the hyperon efficiency was estimated from a comparison 
of the efficiency of the various requirements in a data and Monte Carlo sample. This 
is largest for the \sigp\ and \casz\ modes where the agreement is less satisfactory.
In all modes the uncertainty in the reconstruction is doubled
as there are two hyperons per event.

Table~\ref{tab:syst} displays the breakdown of our 
systematic uncertainties. The last column tabulates the total systematic 
uncertainty added in quadrature for each mode.

\begin{small}
\begin{table*}%[hp]
\begin{center}
\caption{ Breakdown of systematic uncertainty ($\%$) mode by mode. The last column tabulates the 
total systematic uncertainty added in quadrature.}
\label{tab:syst}
\begin{tabular}{ l c c c c c c c c c} \hline
Modes    & $\#\,\psi(2S)$&Bkg. Sub.&Trigger&Tracking&Particle ID&Hyperon    &\piz / \phot&Ang. Dist. & Total    \\ \hline
\ppb     & 3               &-&1      &2$\times$ 1.0   &2$\times$ 1        &  -        &  -         & 3.3       & 4.2          \\
\lzlzb   & 3               &1&1      &4$\times$ 1.0   &2$\times$ 2        & 2$\times$ 1.0     &  -         & 3.3       & 7.7          \\
\spspb   & 3               &-&1      &2$\times$ 1.0   &2$\times$ 2        &2$\times$ 12.8     &  2$\times$ 2       & 3.3       & 26.5  \\
\szszb   & 3               &-&1      &4$\times$ 1.0   &2$\times$ 2        &2$\times$ 1.5      &  2$\times$ 1       & 3.3       & 8.1   \\
\xmxmb   & 3               &-&1      &6$\times$ 1.0   &2$\times$ 2        & 2$\times$ 1.5     &  -         & 3.3       & 9.0          \\
\xzxzb   & 3               &-&1      &4$\times$ 1.0   &2$\times$ 2        &2$\times$ 10.5     &  -         & 3.3       & 22.2         \\
\xsxsb   & 3               &-&1      &8$\times$ 1.0   &2$\times$ 2        &2$\times$ 2.0      &  -         & 10        & 14.4         \\ 
\omomb   & 3               &-&1      &6$\times$ 1.0   &2$\times$ 2+2$\times$ 2    &2$\times$ 2.0      &  -         & 10        & 15.0         \\ \hline
\end{tabular}
\end{center}
\end{table*}
\end{small}

In Table~\ref{tab:result}, we show the measured branching fractions of $\psi(3686)$
decays to baryon-antibaryon modes. 
The branching fractions for octet baryons are in the range of
0.02 - 0.035~$\%$.
Our measured branching fractions for hyperon modes are
$\sim$50$\%$ higher than those in the PDG~\cite{pdg} 
which are based on lower statistics (12 and 8 events in the \casmi\ and \sigz\ modes,  
respectively). We note that isospin partners, \sigp\ and \sigz\ and also
\casmi\ and \casz, have similar branching ratios in agreement with naive expectations.
Furthermore, we note that addition of strangeness does not greatly change the 
branching fraction, demonstrating the flavor symmetric nature of gluons. 
The eighth column in Table~\ref{tab:result}
shows the ratio of these $\psi(2S)$ results to those from $J/\psi$ measurements~\cite{pdg}. 
Two of the results follow the 12$\%$ 
rule closely, but two of them differ by approximately a factor of two. The last column in 
Table~\ref{tab:result} shows the background subtracted continuum cross-section at
$\sqrt{s}\,=\,3.67$ GeV. We quote upper limits $@$90CL for the \lzlzb, \xmxmb, and 
\xzxzb\ production in continuum, owing to marginal signal.
A 20$\%$ correction (upward) is included to account for the initial state radiation~\cite{isr}.

In conclusion, we have analyzed  
CLEO\,III and CLEO-c $\psi(3686)$ data corresponding to
3.08$\times 10^6$ $\psi(3686)$ decays.
We have presented the first observation of the $\psi(3686)$ decaying to \xzxzb\
and \spspb\ final states, and give improved (high statistics) branching ratios
for the $\psi(3686)$ decaying to \ppb, \lzlzb, \xmxmb, and \szszb\ modes.
We also give new upper limits on \xsxsb\ and \omomb\ final states. 

We gratefully acknowledge the effort of the CESR staff 
in providing us with excellent luminosity and running conditions.
This work was supported by the National Science Foundation
and the U.S. Department of Energy.

%\clearpage
%%%%%%%%%%%%%%%%%%%%%%%%%%%%%%%%%%%%%%%%%%%%%%%%%%%%%%%%%%%%%%%%%%%%%%%%%%%% 
 
%%%%%%%%%%%%%%%%%%%%%%%%%%%%%%%%%%%%%%%%%%%%%%%%%%%%%%%%%%%%%%%%%%%%%%%%%%%%%
\end{document}